\documentclass[
    ,final            
  ]
  {aipproc}

\layoutstyle{6x9}

\begin{document}

\title{Charmonium from CLEO}


\author{Kamal K. Seth}{
  address={Department of Physics and Astronomy, 
        Northwestern University, Evanston, IL, 60208, USA}
}

\begin{abstract}
At CLEO, the charmonium singlet states $\eta_c(2^1S_0)$ and $h_c(1^1P_1)$ have been firmly identified and a long standing discrepancy for $\Gamma_{\gamma\gamma}(\chi_{c2})$ has been resolved.
\end{abstract}

\maketitle


\section{1. Introduction}

Quantum Chromodynamics, QCD, took foothold with the discovery of $J/\psi$, the spin--triplet S--wave state $(1^3S_1)$ of Charmonium. Despite 25 years of extensive work in charmonium spectroscopy by SLAC and DESY in $e^+e^-$ annihilation and Fermilab in $p\bar{p}$ annihilation, several nagging problems remained unsolved.  Among the most important were problems related to spin--singlet states and two--photon widths of $C+$ states.

Neither the $e^+e^-$ or $p\bar{p}$ experiments were able to identify the S- and P-wave singlet states $\eta_c(2^1S_0)$ and $h_c(1^1P_1)$.  These states are obviously extremely important for understanding the spin--spin hyperfine interaction of $q\bar{q}$.

The two--photon widths of $C+$ states $\chi_{cJ}(^3P_J:~0^{++},~2^{++})$ are important for understanding relativistic and radiative effects in charmonia, because in the lowest order they are pure QED widths, akin to those of positronium levels.  Unfortunately, results from different measuring techniques have remained very divergent for about twenty years.

In this talk, I am going to address both of these problems.

\section{2. Discovery of $\lowercase{h_c}(1^1P_1)$ State of Charmonium}

After 25 years of failed efforts to identify it, $\eta_c(2^1S_0)$ was finally identified by Belle, BaBar, and CLEO.  CLEO \cite{cleo-etacp} measured $M(\eta_c(2^1S_0))=3642.9\pm3.4$ MeV, which determines $\Delta M_{hf}(2S)=43.1\pm3.4$ MeV.  This is an unexpected and extremely interesting result, considering that $\Delta M_{hf}(1S)=117\pm1$ MeV, and most theoretical predictions had $\Delta M_{hf}(2S)\approx 65$ MeV.

The unexpectedly small hyperfine splitting observed for 2S states makes it more interesting than ever to look deeper into the hyperfine splitting of other charmonia, in particular the 1P states.

If the confinement potential is Lorentz scalar, there is no long--range spin--spin interaction in $q\bar{q}$.  It follows that for all other waves ($L\ne0)$ hyperfine splitting is zero, so that $\Delta M_{hf}(1P)=M(\left<^3P_J\right>)-M(^1P_1)$. To test this prediction it is necessary to identify $h_c(1^1P_1)$ and measure $M(h_c)$ with precision.

In 1992 Fermilab E760 studied the reaction $p\bar{p}\to h_c \to\pi^0 J/\psi$ and claimed the observation of a signal for $h_c$.  However, higher luminosity runs by Fermilab E835 \cite{e835-hc} in 1996 and 2000 failed to confirm this observation.  Fermilab E835 has also searched for $h_c$ in their 1996/2000 data in the reaction $p\bar{p}\to h_c \to\gamma\eta_c.$ They report $\Delta M_{hf}(1P)=-0.4\pm0.2\pm0.2\;\mathrm{MeV}$, with 13 observed events and a significance of the $h_c$ signal at $\sim3\sigma$ level. At CLEO \cite{cleo-hc} we have now firmly identified $h_c$ with a significance of $6\sigma$.

At CLEO data were taken at $\psi(2S)$, with 3.08 million $\psi(2S)$.  These data have been analyzed for \cite{cleo-hc}
$$\psi(2S)\to\pi^0h_c\;,\; h_c\to\gamma\eta_c$$
Both inclusive and exclusive analyses were done, and an accurate determination of $h_c$ mass was made in recoils against $\pi^0$'s whose energy could be measured with precision.

\textbf{Inclusive Analyses:} Two independent analyses were made, one in which the photon energy $E_\gamma$ was constrained, and the other in which $M(\eta_c)$ was constrained.  Completely consistent results were obtained.

\textbf{Exclusive Analysis:} In this analysis, instead of constraining $E_\gamma$ or $M(\eta_c)$, seven known decay channels of $\eta_c$ were measured.  Once again, consistent results were obtained.

The overall result is $M(h_c) = 3524.4\pm0.6\pm0.4\;\mathrm{MeV},$ or
$$\Delta M_{hf}(1P)=\left<M(\chi_{cJ})\right>-M(h_c)=+1.0\pm0.6\pm0.4\;\mathrm{MeV}$$

Two conclusions follow from these results: (a) the simple pQCD expectation is not strongly violated, (b) the magnitude and sign of $\Delta M_{hf}$ is not yet well determined.

\begin{figure}[!tb]
\includegraphics[width=2.2in]{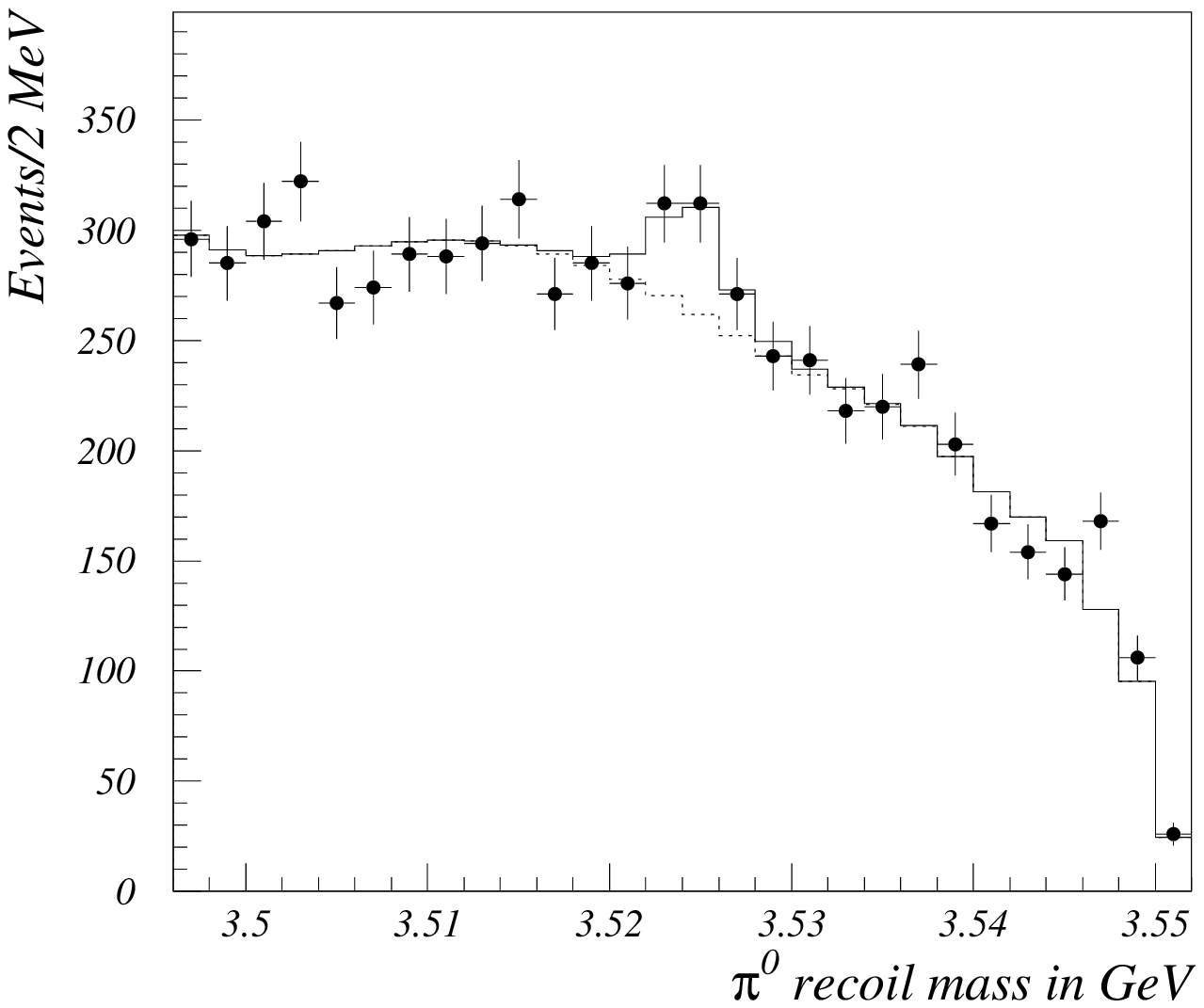}
\raisebox{1.7in}{\rotatebox{270}{\includegraphics[width=1.5in]{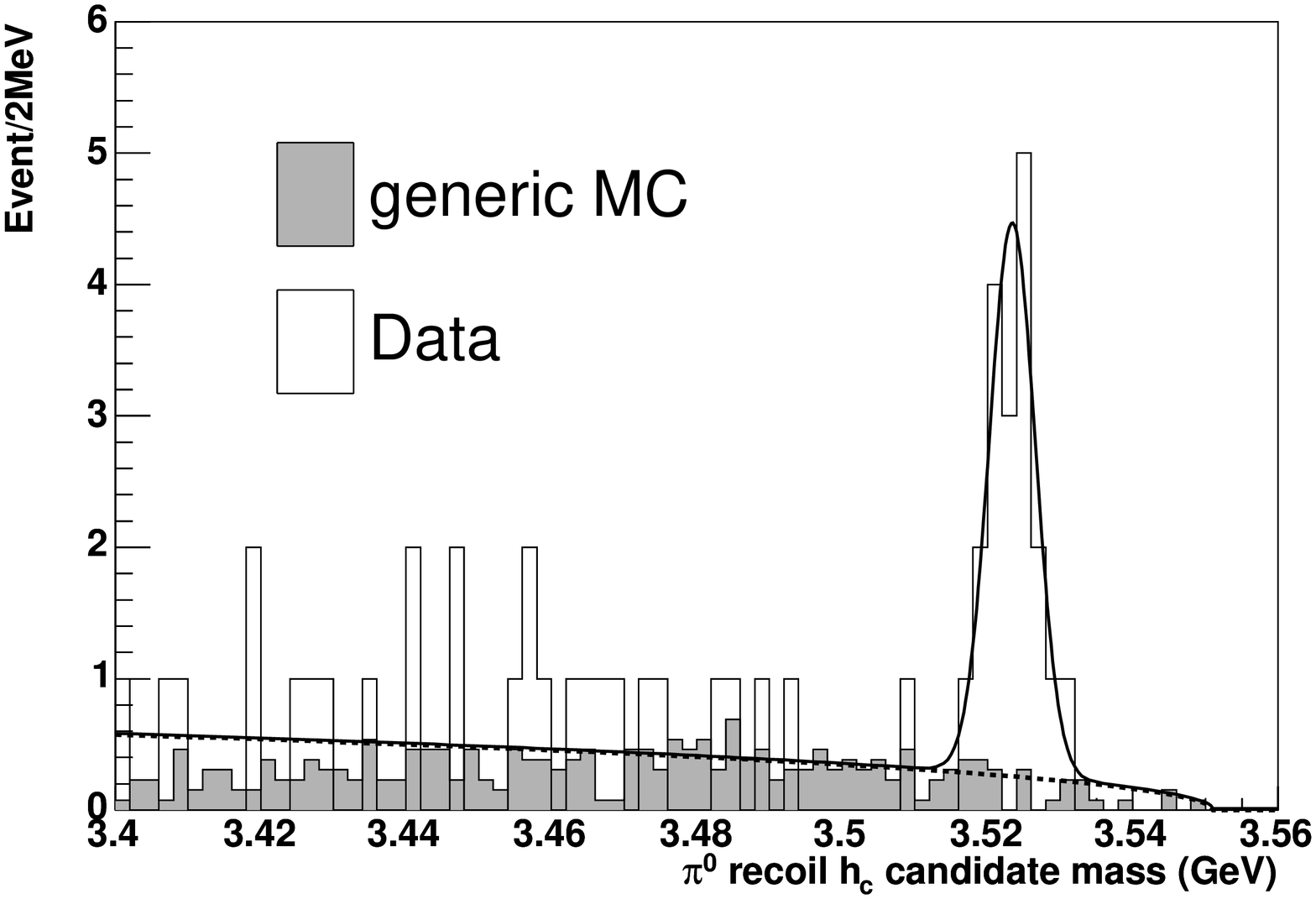}}}
\caption{CLEO Observation of $h_c(1^1P_1)$, in (left) inclusive analysis (right) exclusive analysis \cite{cleo-hc}.}
\end{figure}

\section{3. Two--Photon Width of $\chi_{\lowercase{c}2}$}

There are two different ways of measuring the two--photon width of $\chi_{c2}$.  In $e^+e^-$ collisions, the $\chi_{c2}$ state is formed in two--photon fusion, $\gamma\gamma\to\chi_{c2}$, and a subsequent decay of $\chi_{c2}$ (usually $\chi_{c2}\to\gamma J/\psi$) is measured.  These measurements, including the latest one from Belle \cite{belle-chi2gam}, yield $\Gamma_{\gamma\gamma}(\chi_{c2})=1000-3000$ eV.  In $p\bar{p}$ annihilation, the $\chi_{c2}$ state is directly formed and its decay into two--photons is measured. In these Fermilab measurements \cite{e835-chi2gam}, $\Gamma_{\gamma\gamma}(\chi_{c2})\approx300$ eV is determined.  It is this persistent and long-standing large discrepancy which motivated us to make the present measurement at CLEO \cite{cleo-chi2gam} using 14.4 fb$^{-1}$ of $e^+e^-$ data taken in the Upsilon region to study the two--photon fusion reaction
$$e^+e^-\to e^+e^-+\gamma\gamma,~\gamma\gamma\to\chi_{c2}\to\gamma J/\psi,~J/\psi\to e^+e^-+\mu^+\mu^-$$

The spectrum of photon mass, $\Delta M \equiv M(\gamma l^+l^-) - M(l^+l^-)$, in Fig. 3 shows clear peaks corresponding to the E1 photon from $\chi_{c2}\to\gamma J/\psi$.  Fits to these peaks leads to $\Gamma_{\gamma\gamma}(\chi_{c2})=544\pm87$, $571\pm76$, $559\pm57$ eV for $e^+e^-$, $\mu^+\mu^-$, and ($e^+e^-+\mu^+\mu^-$), respectively, using $\mathcal{B}(\chi_{c2}\!\to\!\gamma J/\psi)\!=\!(19.9\pm1.7)\%$ as measured recently by CLEO \cite{cleo-psirad}.

\begin{figure}[!tb]
\begin{tabular}{ll}
\includegraphics[width=1.6in]{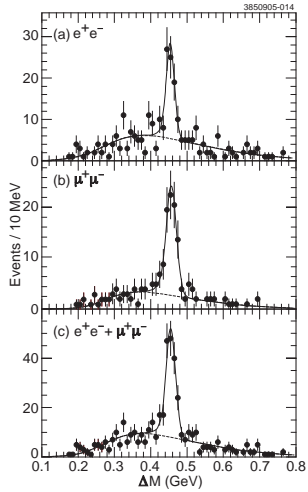}
&
\raisebox{1.4in}{\parbox{4.in}{
{\small \textbf{TABLE 1.}~~~Results of the latest measurements of $\Gamma_{\gamma\gamma}(\chi_{c2})$.}\\

\begin{footnotesize}
\begin{tabular}{ccc}
\hline \hline
Experiment & $\Gamma_{\gamma\gamma}(\chi_{c2})$ (eV) & $\Gamma_{\gamma\gamma}(
\chi_{c2})$ (eV) \\
 & (as published) & (as reevaluated) \\ \hline
Present \cite{cleo-chi2gam}: $\gamma\gamma\to\chi_{c2}$ & \multicolumn{2}{c}{$559\pm83$} \\
\hline
Belle \cite{belle-chi2gam}:  $\gamma\gamma\to\chi_{c2}$ & $850\pm127$ &
$570\pm81$ \\ \hline
E835 \cite{e835-chi2gam}: $\chi_{c2}\to\gamma\gamma$ &  $270\pm59$ & $384\pm83$ \\
 \hline \hline
\end{tabular}
\end{footnotesize}}}
\\
\end{tabular}
\caption{Distributions of $\Delta M$ observed. \hspace*{3.in}}

\end{figure}


In making a comparison of the present results with other recent results, it was noted that the differences originated from using an old (PDG1990) value for $\mathcal{B}(\chi_{c2}\to\gamma J/\psi)$ which was nearly 40\% smaller than the result of the new CLEO measurement \cite{cleo-psirad}.  When the results of the Belle and E835 measurements are reevaluated using the CLEO value of  $\mathcal{B}(\chi_{c2}\to\gamma J/\psi)$, it is found that all results become completely consistent.

Since the two gluon decay width of $\chi_{c2}$, $\Gamma_{gg}(\chi_{c2})=1.55\pm0.11$ MeV, our measurement of $\Gamma_{\gamma\gamma}(\chi_{c2})$ allows us to estimate the strong coupling constant $\alpha_S(m_c)$.  According to pQCD, ${\Gamma_{\gamma\gamma}(\chi_{c2})/\Gamma_{gg}(\chi_{c2})} = ({8\alpha^{2}/ 9\alpha_{s}^{2}})$.  This leads to $\alpha_S=0.36\pm0.03$.  If first order radiative corrections are included, we obtain $\alpha_S=0.29\pm0.03$.






\bibliographystyle{aipproc}   

\begin{thebibliography}{99}



\bibitem{cleo-etacp} CLEO Collaboration, Phys. Rev. Lett. \textbf{92} (2004) 142001, and references therein.

\bibitem{e835-hc} Fermilab E835 Collaboration, Phys. Rev. \textbf{D 72} (2005) 032001.

\bibitem{cleo-hc} CLEO Collaboration, Phys. Rev. Lett. \textbf{95} (2005) 102003.

\bibitem{belle-chi2gam} Belle Collaboration, Phys. Lett. \textbf{B 540} (2002) 33.

\bibitem{e835-chi2gam} Fermilab E835 Collaboration, Phys. Rev. \textbf{D 62} (2000) 052002.

\bibitem{cleo-chi2gam} CLEO Collaboration, hep-ex/0510033, submitted to Phys. Rev. Lett.

\bibitem{cleo-psirad} CLEO Collaboration, Phys. Rev. \textbf{D 70} (2004) 112002.

\end{thebibliography}


\end{document}